\documentstyle[twocolumn,aps,psfig]{revtex}
\tighten
\begin{document}
\draft

\title{
Role of Spatial Amplitude Fluctuations in 
Highly Disordered s-Wave Superconductors
}

\author{
Amit Ghosal, Mohit Randeria and Nandini Trivedi
}

\address{ Department of Theoretical Physics,
Tata Institute of Fundamental Research, Mumbai 400005, India \\
}
\address{
\begin{minipage}[t]{6.0in}
\begin{abstract}
The effect of non-magnetic impurities on 2D s-wave superconductors is studied
beyond the weak disorder regime.  
Within the Bogoliubov-de Gennes (BdG) framework,
the local pairing amplitude develops a broad distribution
with significant weight near zero with increasing disorder.
Surprisingly, the density of states continues to show a finite spectral gap.
The persistence of the spectral gap
at large disorder is shown to arise from the break up of the system into
superconducting ``islands".
Superfluid density and off-diagonal correlations show
a substantial reduction at high disorder.
A simple analysis of phase fluctuations about the highly inhomogeneous 
BdG state is shown to lead to a transition to a non-superconducting state.
\end{abstract}
\pacs{PACS numbers: 74.20.-z, 74.40.+k, 74.20.M}
\end{minipage}}
\date{\today}
\maketitle

\narrowtext
The effect of strong disorder on superconductivity has been a subject of
considerable interest, both theoretically \cite{rama,th_reviews} 
and experimentally \cite{ex_review,exp}, for a long time.
A generally accepted physical picture of how the superconducting
(SC) state is destroyed and the nature of the non-SC state has
not yet emerged.
Much of the theoretical work
(``pairing of exact eigenstates'' \cite{rama,exact_eig} or 
diagrammatics \cite{th_reviews,diagrams}) assumes
that the pairing amplitude $\Delta({\bf r})$ is 
uniform in space (${\bf r}$-independent) even for a highly disordered SC; see
however \cite{sad,franz}.
Recent work on universal properties at the SC-insulator transition
\cite{dirty_bosons}
has also ignored amplitude fluctuations, since phase fluctuations are
presumably responsible for critical properties.

In this paper we consider a simple model of a 2D s-wave superconductor 
at $T = 0$
in a random potential, defined by eqn.~(1) below, and analyze it in detail
within a Bogoliubov-deGennes (BdG) framework \cite{degennes}. 
Our goal is to see how the
local pairing amplitude $\Delta({\bf r})$ varies spatially in the presence
of disorder, and the effect of this inhomogeneity on physically relevant
correlation functions. Our results can be summarized as follows:

\noindent (1) With increasing disorder, the distribution $P(\Delta)$ 
of the local paring amplitude $\Delta({\bf r})$ becomes very broad,
 eventually developing considerable weight near $\Delta\approx 0$.

\noindent (2) 
 The spectral gap in the one-particle
density of states persists even at high disorder in spite of the 
growing number of sites with $\Delta({\bf r})\approx 0$. 
A detailed understanding of this
surprising effect emerges from a study of the spatial variation of
$\Delta({\bf r})$ and of the BdG eigenfunctions.

\noindent (3) There is substantial reduction in the superfluid stiffness and
off-diagonal correlations with increasing disorder, however,
the amplitude fluctuations by themselves cannot destroy the superconductivity.

\noindent (4) Phase fluctuations about the inhomogeneous 
BdG state are described by a quantum XY model whose parameters, 
compressibility and phase stiffness, are obtained from the BdG results. 
A simple analysis of this effective model within a self-consistent harmonic 
approximation leads to a transition to a non-SC state. 

We conclude with some comments on the implications of our results for 
experiments on disordered films.


We model the 2D disordered s-wave SC by an attractive Hubbard model
with on-site disorder:
\begin{equation}
{\cal H} = {\cal K} - |U|\sum_{i} n_{i \uparrow} n_{i \downarrow} 
 + \sum_{i,\sigma} (V_{i}-\mu ) n_{i\sigma}.
\label {eq:hamil}
\end{equation}
${\cal K} = -t\sum_{<ij>,\sigma} (c_{i\sigma}^{\dag} c_{j\sigma} + h.c.)$
is the kinetic energy,
$c_{i\sigma}^{\dag}$ ($c_{i\sigma}$) the creation (destruction) 
operator for an electron with spin $\sigma$ on a site ${\bf r}_i$ of
a square lattice,
$t$ the near-neighbor hopping, 
$|U|$ the pairing interaction,
$n_{i\sigma}= c_{i\sigma}^{\dag}c_{i\sigma}$, and
$\mu$ the chemical potential.
The random potential $V_{i}$ is chosen independently 
at each ${\bf r}_i$ from a uniform distribution $[-V,V]$;
$V$ thus controls the strength of the disorder. 

We begin by treating the spatial fluctuations of the pairing amplitude
using the standard BdG equations \cite{degennes}: 
\begin{equation}
\left(\matrix{\hat\xi & \hat\Delta \cr \hat\Delta^{*} & -\hat\xi^{*}} \right)
\left(\matrix{u_{n}({\bf r}_i) \cr v_{n}({\bf r}_i)} \right) = E_{n}
\left(\matrix{u_{n}({\bf r}_i) \cr v_{n}({\bf r}_i)} \right)
\label {eq:bdg}
\end{equation}
where
$\hat\xi u_{n}(r_i) = -t\sum_{\hat\delta}
u_{n}({\bf r}_i+\hat\delta)+(V_i-\tilde{\mu_i})u_{n}({\bf r}_i)$
and $\hat\Delta u_{n}({\bf r}_i) = \Delta({\bf r}_i) u_{n}({\bf r}_i)$,
and similarly for $v_{n}({\bf r}_i)$. 
Here $\hat\delta = \pm{\hat{\bf x}}, \pm{\hat{\bf y}}$ and
$\tilde{\mu_i} = \mu + |U|n_i/2$ incorporates the site-dependent 
Hartree shift in presence of disorder.
Starting with an initial guess for $\Delta({\bf r}_i)$'s and $\tilde{\mu_i}$
we numerically solve for the BdG eigenvalues $E_n$ and eigenvectors 
$\left(u_{n}({\bf r}_i),v_{n}({\bf r}_i)\right)$
on a finite lattice of $N$ sites with periodic boundary conditions. 
We then calculate
the local pairing amplitudes and number density at $T=0$, given by
\begin{eqnarray}
\Delta({\bf r}_i) = |U|\sum_{n}u_{n}({\bf r}_{i})v_{n}^{*}({\bf r}_{i}),~~~
n_i = \sum_{n}|v_{n}({\bf r}_{i})|^{2}
\label {eq:selfc}
\end{eqnarray}
and iterate the process until self-consistency is achieved 
for $n_i$ and $\Delta({\bf r}_i)$ {\em at each site}. $\mu$ is 
determined by $1/N \sum_i n_i = \langle n \rangle$, 
where $\langle n \rangle$
is the average density.

We have studied the model (1) for a range of parameters:
$1 \leq |U|/t \leq 8$ and $0 \leq V/t \leq 12$ on lattices of size 
$N = 12 \times 12$ (some checks were made on $24 \times 24$ systems). 
We focus below on $|U|/t = 2$ and $4$, $\langle n \rangle=0.875$;
similar results are obtained for other parameters. 
The number of iterations necessary to obtain self-consistency grows
with disorder; we have checked that the same
solution is obtained for different
initial guesses. Results are averaged 
over 16-20 different realizations of the disorder.


The distribution $P(\Delta)$ of local pairing amplitudes for $|U|=4$
is plotted in Fig.~1. 
For $V\stackrel {_<}{_\sim} 0.25t$,
$P(\Delta)$ has a sharp peak near the $V=0$ BCS value of 
$\Delta_0 \simeq 1.36t$.
In the small $V$ limit, pairing of exact eigenstates is justified,
since this naturally leads \cite{uniform}
to uniform $\Delta({\bf r})$. However, this approximation fails
with increasing $V$ as $P(\Delta)$ becomes
extremely broad for $V\sim t$, eventually becoming
rather skewed at $V\geq 2t$ with 
a large number of sites with $\Delta({\bf r})\approx 0$.


\begin{figure}
\vskip-2mm
\hspace*{0mm}
\psfig{file=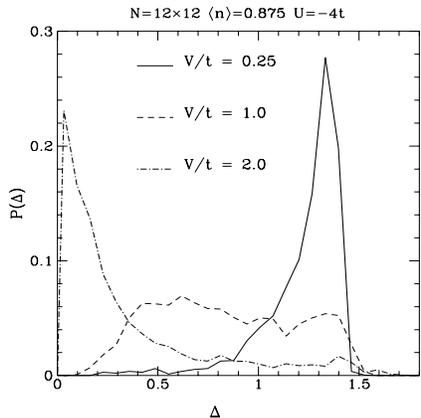,height=2.2in,width=2.3in,angle=0}
\vskip0mm
\caption{
The distribution $P(\Delta)$ of the pairing amplitude 
$\Delta({\bf r}_i)$ for different disorder strengths $V$.
For small $V$, $P(\Delta)$ is peaked around the BCS 
$\Delta_0$, but becomes increasingly broad at higher $V$, indicative
of a highly inhomogeneous state. 
}
\label{fig:P_delta}
\end{figure}


\begin{figure}
\vskip-5mm
\hspace*{0mm}
\psfig{file=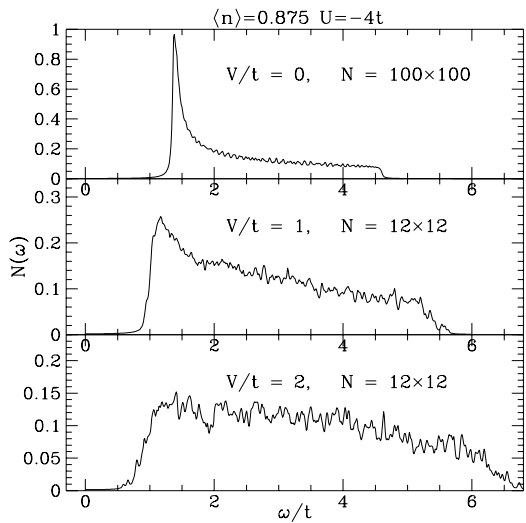,height=2.8in,width=2.8in,angle=0}
\vskip0mm
\caption{
Density of states $N(\omega)$ for three disorder strengths $V$ which show the
the persistence of a spectral gap at all disorder;
(Note the different vertical scale for each case).
}
\label{fig:dos}
\end{figure}
To study how the spectral gap evolves as the pairing amplitude
becomes highly inhomogeneous, we look at the (disorder averaged)
one-particle density of states (DOS) 
$N(\omega) = 1/N \sum_n \delta(\omega - E_n)$,
defined in terms of the BdG eigenvalues $E_n$.
(Numerically, $\delta$-functions 
are broadened into Lorentzians with a width
of order spacing between $E_n$'s).
From Fig.~2 we see that with increasing disorder
the DOS pile-up at the gap edge is progressively smeared out,
and that states are pushed up to higher energies.
But the most remarkable feature of Fig.~2
is the presence of a finite spectral gap even at high disorder. 
While we can not rule out
an exponentially small tail in the low energy DOS from a finite
system calculation, we always found, for each disorder realization,
that the lowest BdG eigenvalue remains non-zero and of the order of
the zero-disorder BCS gap; see also Fig.~3(a).
We also emphasize that approximate treatments of the BdG equations
\cite{wheatley}, which do not treat the local amplitude fluctuations
properly, miss this remarkable feature, as do simplified models in
which $\Delta({\bf r}_i)$'s are assumed to be independent random variables
at each site.


\begin{figure}
\vskip-2mm
\hspace*{0mm}
\psfig{file=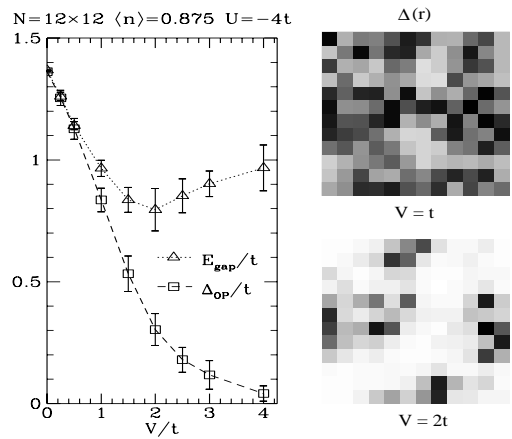,height=2.4in,width=2.7in,angle=0}
\vskip0mm
\caption{
(a) Left panel:
$T= 0$ spectral gap $E_{\rm gap}$
and order parameter $\Delta_{\rm OP}$ (see text)
as functions of disorder $V$.
The two coincide for small $V$ but become very different at large disorder.
(b) Right Panel:
Gray-scale plot showing the spatial variation of
$\Delta({\bf r}_i)$ for the same disorder configuration
with different $V$.
Larger $\Delta({\bf r}_i)$'s are indicated by darker shades.
Note the spatially correlated structures at $V=2t$ with
``SC islands'' separated by a ``sea'' of nearly vanishing $\Delta$'s.
}
\label{fig:odlro_gap}
\end{figure}
To understand the persistence of a finite spectral gap at high
disorder, when a large fraction of the sites have near vanishing
pairing amplitude, it is useful to study the spatial variation of 
of the $\Delta({\bf r}_i)$'s 
and the BdG eigenvectors $\left(u_{n}({\bf r}_i),v_{n}({\bf r}_i)\right)$
{\em for individual realizations of the disorder potential}.
A particularly simple picture emerges at high disorder:
there are spatially correlated clusters of sites at which  
$\Delta({\bf r}_i)$ is large (``SC islands''), and these are separated by 
large regions where $\Delta({\bf r}_i) \approx 0$ (see Fig.~3(b)).
We find that the SC islands correlate well with regions where the
absolute magnitude of the random potential $|V_i|$ is small; deep valleys and 
high mountains in the potential do not allow for number fluctuations and
are thus not conducive to pairing. The density $n({\bf r})$ is
also highly inhomogeneous, and
for moderate $|U|$ ($\geq 4t$) and high disorder, we have found clear evidence
for ``particle-hole mixing in real space'', i.e., a spatial correlation
between $\Delta({\bf r})$ and $n({\bf r})/2[1 - n({\bf r})/2]$ 
\cite{long_paper}.

At high disorder, we found that the eigenfunctions corresponding to 
low-lying excitations live entirely on the SC islands (i.e., the darker 
regions in Fig.~3 (b))
resulting in the finite spectral gap. 
On the other hand, regions where the pairing amplitude is small
correspond to very large values of $|V_i|$, as explained above,
and thus support even higher energy excitations. 
Clearly this simple picture of SC islands is well defined only in
the large disorder regime, nevertheless, it is 
useful for understanding
the spectral gap in this limit. In the opposite limit of low disorder,
of course, the BCS-like spectral gap is obvious.

We next turn to the question of how superconductivity is affected
in the highly inhomogeneous BdG state. The off-diagonal long range order
parameter $\Delta_{\rm OP}$ is defined 
by the (disorder averaged) correlation function
$\langle c_{i\uparrow}^{\dag}c_{i\downarrow}^{\dag}c_{j\downarrow}
c_{j\uparrow} \rangle \rightarrow \Delta_{\rm OP}^2/|U|^2$
for large $|{\bf r}_i-{\bf r}_j|$.
From Fig.~3 (a) we see
that $\Delta_{\rm OP}$ is the same as the spectral gap
(and both equal the uniform pairing amplitude) 
for small disorder, as expected from BCS theory.
However beyond a certain $V$ the two quantities deviate from each other:
in contrast to the spectral gap, the order parameter decreases
with increasing disorder; (we find that $\Delta_{\rm OP} \simeq 
\int d\Delta \Delta P(\Delta)$, i.e., the average value of the
pairing amplitude).

The superfluid stiffness $D_s^0$ 
 is given by\cite{swz}
$D_{s}^0/\pi = \langle -k_{x} \rangle -
\Lambda_{xx}(q_{x}=0,q_y \rightarrow 0,\omega=0)$.
The diamagnetic term $\langle -k_x\rangle$, 
is one-half (in 2D) the kinetic energy 
$\langle-{\cal K}\rangle$, and the paramagnetic 
term $\Lambda_{xx}$ is the (disorder
averaged) transverse current-current correlation function. We have also
checked that the charge stiffness $D^0$ is equal to $D_s^0$.
($D^0$ is the strength of the delta-function in $\sigma(\omega)$,
and given in terms of $\Lambda_{xx}({\bf q}=0,\omega\rightarrow 0)$
 \cite{swz}). 

\begin{figure}
\vskip-3mm
\hspace*{0mm}
\psfig{file=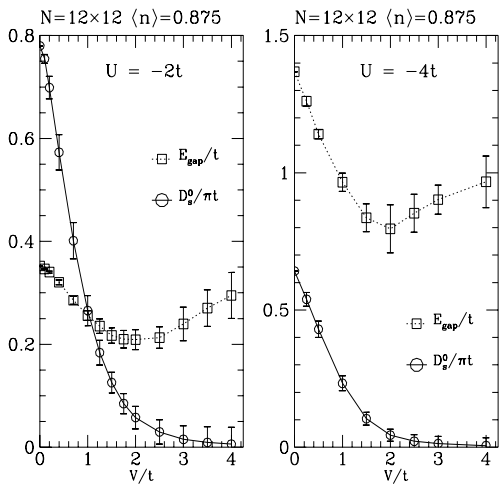,height=2.7in,width=2.7in,angle=0}
\vskip2mm
\caption{
The $T=0$ BdG superfluid stiffness $D_s^0$ and the spectral gap $E_{\rm gap}$
plotted as a function of disorder strength for two different
values of attraction $|U|$.
Note that irrespective of whether $D_s^0$ is larger than or comparable
to $E_{\rm gap}$ at $V=0$, the gap persists with increasing disorder while
stiffness decreases.
}
\label{fig:sd_gap}
\end{figure}

The $D_s^0$ calculated within BdG theory
shows a large reduction \cite{inhom} by two orders of magnitude
with increasing disorder; see Fig.~4.
We see that for $U= -2t$, at $V=0$, 
$D_s^0 \gg E_{\rm gap}$, characteristic of weak coupling BCS theory,
where the vanishing of the gap determines $T_c$,
while for $U = -4t$, $D_s^0$ and $E_{\rm gap}$ are comparable
at $V=0$, indicative of an intermediate coupling regime \cite{rtms}
where thermal phase fluctuations are important for
determining $T_c$ \cite{emery}.
However, for {\em all} $|U|/t$, we always find $D_s^0 \ll E_{\rm gap}$
at large disorder, and thus phase fluctuations have to be taken into account.
In fact the reason why $D_s^0$ is not driven to zero
at large $V$ within the BdG framework is due to the
neglect of these fluctuations.

To make a rough estimate of the effect of phase fluctuations 
about the inhomogeneous BdG state we use a quantum XY model with
an effective Hamiltonian \cite{rama}
$H_{\theta} = -(\kappa/8)\sum_j \dot{\theta}_j^2 + 
(D_s^0/4)\sum_{\langle j k \rangle } \cos(\theta_j - \theta_k)$,
whose parameters are obtained from the preceding analysis:
the bare $D_s^0$ is the BdG superfluid stiffness and 
$\kappa=dn/d\mu$ is the BdG compressibility.
The large reduction in $dn/d\mu$
with disorder seen in Fig.~5 (a) can be understood 
qualitatively at large $V$
in terms of the charging energy of the SC islands.
Note that, in this simplified description using $H_{\theta}$, we ignore 
the inhomogeneity in the local bare stiffness and charging energies.
\begin{figure}
\vskip-8mm
\hspace*{0mm}
\psfig{file=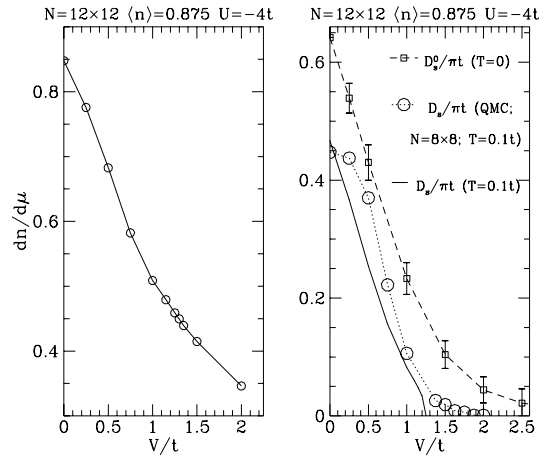,height=2.7in,width=2.8in,angle=0}
\vskip-2mm
\caption{
(a) Left Panel: The compressibility $dn/d\mu$ as a function of disorder $V$.
(b) Right Panel: The superfluid density $D_{s}/\pi$ as a function of
disorder $V$. By including phase fluctuation a transition is obtained at
$V_c$ which is very close to that predicted by QMC \protect \cite{qmc}.
}
\label{fig:phase}
\end{figure}


We use a variational approximation \cite{phase} 
to estimate the renormalized superfluid stiffness 
$D_s = D_s(\kappa,D_s^0)$, by finding the best harmonic  
$H_{\rm trial} = -(\kappa/8)\sum_j \dot{\theta}_j^2 + 
(D_s/8)\sum_{\langle j k \rangle } (\theta_j - \theta_k)^2$,
which describes $H_{\theta}$. 
The phase variables $\theta_i$ are assumed to live on
a lattice with lattice constant set by the BdG coherence length $\xi_0$.
For $U = -4t$ we choose $\xi_0 \simeq 1.8$ \cite{cutoff} 
by demanding that the renormalized $D_s$ at $V=0$ agrees with that
obtained from quantum Monte Carlo (QMC) \cite{qmc} ($D_s/\pi \simeq 0.45t$) 
for the pure case. 

We now calculate the renormalized $D_s$ as a function of disorder,
using the $V$-dependent $\kappa$ and $D_s^0$ from the BdG analysis
as input and keeping $\xi_0$ fixed; 
(details will be presented elsewhere \cite{long_paper}).
As shown in Fig.~5 (b), $D_s$ is driven to
zero beyond a critical disorder $V_c$ which is in very reasonable
agreement with QMC \cite{qmc}.
Thus a transition to a non-SC (insulating) state is indeed obtained
by incorporating the effects of phase fluctuations about the
inhomogeneous BdG state.

We emphasize that the finite spectral gap 
obtained in the BdG analysis at large $V$ will survive inclusion of
phase fluctuations, since this gap is related to the inhomogeneous
SC islands.
A key question is whether the inhomogeneous $\Delta({\bf r})$ leading
to a spectral gap in the insulating state persists all the way
down to $|U|/t \ll 1$. A definitive answer cannot be obtained since
weak coupling BdG calculations are plagued by severe finite size effects
\cite{rts}. 
It is important to note that the gap persists in
the $|U|=2t$ case (see Fig.~4(a)) which in the $V=0$ limit 
has $D_s^0 \gg E_{\rm gap}$,
characteristic of weak coupling SC. The available numerical results
suggest that even for weak coupling, $\Delta({\bf r})$ inhomogeneities
are generated on the scale of the coherence length, which eventually
show up as SC islands at large disorder. This would suggest persistence 
of the gap.
In contrast, some tunneling experiments \cite{exp}(c) show 
a finite DOS N(0) with increasing disorder, which then points to
physical effects beyond those in the simplest model studied here.
One possibility is that Coulomb interactions in
the presence of disorder lead to an effective $|U|$ which is
${\bf r}$-dependent, and regions with $|U_i| \approx 0$ lead to
finite $N(0)$. Another possibility is that Coulomb interactions plus disorder
lead to the formation of local moments which are pair breaking.

Another implication of our results for experiments is that
SC-Insulator transitions in disordered films are often described in terms
of two different paradigms: homogeneously disordered films (driven insulating
by the vanishing of the gap) and granular
films (driven by vanishing of the phase stiffness). 
In our simple model, although the system was homogeneously disordered
at the microscopic level, 
granular SC-like  structures developed in so far as
the pairing amplitude was concerned. 
It would be very interesting to use
STM measurements to study variations in the local density states to shed more
light on this question. 

{\bf Acknowledgements}: We would like to thank A. Paramekanti,
T. V. Ramakrishnan and R. T. Scalettar for useful discussions.



\begin{references}
\bibitem{cutoff}
The renormalized $D_s$ is highly sensitive to the choice of
$\xi_0$. ( e.g., in weak coupling BCS theory with $\xi_0 \gg 1$, one has
$D_s=D_s^0$). Our choice of $\xi_0$, by comparing $D_s$ with QMC,
is also consistent with independent estimates from the BdG analysis.
\end{references}
\end{document}